# Generation of degenerate, factorizable, pulsed squeezed light at telecom wavelengths

Thomas Gerrits[1], Martin J. Stevens[1], Burm Baek[1], Brice Calkins[1], Adriana Lita[1], Scott Glancy[1], Emanuel Knill[1], Sae Woo Nam[1], Richard P. Mirin[1], Robert H. Hadfield[2], Ryan S. Bennink[3], Warren P. Grice[3], Sander Dorenbos[4], Tony Zijlstra[4], Teun Klapwijk[4], Val Zwiller[4]

[1]National Institute of Standards and Technology, 325 Broadway, Boulder, CO 80305, USA
[2]Scottish Universities Physics Alliance and School of Engineering and Physical Sciences, Heriot-Watt University, Edinburgh, EH14 4AS, UK
[3]Center for Quantum Information Science, Oak Ridge National Laboratory, Oak Ridge, TN, 37831, USA
[4]Kavli Institute for Nanoscience, Delft University of Technology, Lorentzweg 1, 2628 CE Delft, The Netherlands
*gerrits@boulder.nist.gov

**Abstract:** We characterize a periodically poled KTP crystal that produces an entangled, two-mode, squeezed state with orthogonal polarizations, nearly identical, factorizable frequency modes, and few photons in unwanted frequency modes. We focus the pump beam to create a nearly circular joint spectral probability distribution between the two modes. After disentangling the two modes, we observe Hong-Ou-Mandel interference with a raw (background corrected) visibility of 86 % (95 %) when an 8.6 nm bandwidth spectral filter is applied. We measure second order photon correlations of the entangled and disentangled squeezed states with both superconducting nanowire single-photon detectors and photon-number-resolving transition-edge sensors. Both methods agree and verify that the detected modes contain the desired photon number distributions.

---

## 1. Introduction

One of the most familiar non-classical states of light is a squeezed vacuum, which is typically created through a nonlinear process that generates correlated pairs of signal and idler photons from an initial pump field [1]. Squeezing is generally characterized with homodyne detection by mixing the squeezed vacuum on a beam splitter with a reference local-oscillator field. A *pure* squeezed vacuum has no excess noise; in other words, the uncertainty product of the squeezed and anti-squeezed quadratures exactly equals the minimum allowed by the uncertainty principle. Many applications of squeezed light, from precision metrology to quantum information, benefit from access to squeezing with high purity.

Homodyne detection can be used to measure the *x*- and *p*-quadratures of the electromagnetic field, and using the variances of these quadratures, one can estimate the purity of a quantum state. The purity is given by $\text{tr}(\rho^2) = 1/\left(2\sqrt{\text{var}(x)\text{var}(p)}\right)$, where var(*x*) and var(*p*) are the respective variances, the vacuum quadrature variance is ½, and $\rho$ is the density matrix [2]. The observed purity can be degraded by losses (such as those at free-space optical components and inefficiencies of the photodetectors), electronic noise, and mode-overlap mismatch of a multi-mode squeezed vacuum with the measurement mode, *i.e.*, the local-oscillator mode. To date, high levels of pure squeezed vacuum have been achieved with cavity squeezing using a CW laser source [3, 4]. However, no such levels of pure squeezed vacuum have yet been demonstrated by use of a pulsed source. A pulsed source of pure squeezed vacuum is especially interesting because it can qualify for directly heralding high-fidelity optical Schrödinger cat states *via* photon subtraction [5-7].

Because homodyne detection is not sensitive to modes that are orthogonal to the local oscillator's mode, photons in these modes have no effect on the observed purity. Nevertheless, photons in unwanted modes can cause significant problems when the potentially multi-mode squeezed source is used for experiments involving photon subtraction, because typical photon counters are sensitive to these photons. In addition to high purity squeezing, these experiments also require that all modes (spatial and spectral-temporal) of all three fields (the squeezed source, the subtracted photons, and the local oscillator) must all overlap one another. For ideal photon subtraction, the nonlinear process should produce squeezed light in a single mode, allowing *all* the subtracted photons to match the mode of the local oscillator and the mode of the single-mode optical fiber used to spatially filter the subtracted photons.

In this work our goal is to create two entangled squeezed modes (signal and idler modes) with orthogonal polarizations and identical spatio-temporal distributions from a periodically poled potassium titanyl phosphate (pp-KTP) crystal, so that their interference at a 50/50 beam splitter creates two single-mode squeezed vacua at the two output ports of that beam splitter. A second goal is to avoid the production of light in other modes emerging from the crystal. We will show how these two goals are met by engineering both the pp-KTP crystal and the optical pulse used to pump this crystal.

To characterize this source, we employ a variety of methods that are sensitive to photons in the unwanted modes and can determine the mode matching of the signal and idler modes from the two entangled squeezed modes before they are combined at a 50/50 beam splitter. These methods can be used to investigate the mechanisms by which the mode matching is degraded, but typically cannot quantify optical losses. By measuring correlations between the wavelengths of the signal and idler photons, one can construct the joint spectral probability distribution. If the two entangled squeezed modes are generated in a factorizable state, we should see that the joint spectral probability distribution is also factorizable [8, 9]. In the low-squeezing limit, where a single photon pair is generated, it was recently shown that it is possible to generate pairs with an effective mode number that is close to the ideal value of 1 [8-15]. Hong-Ou-Mandel interference provides a measure of the similarity of the spectral distributions of the signal and idler modes. The second-order coherence, $g^{(2)}$, is another important metric for characterizing the mode structure of squeezed light: mismatch between the signal and idler modes or photons in unwanted modes will cause the measured $g^{(2)}$ values to depart from theoretical predictions.

In the following sections we discuss measurements of the joint spectral distribution of our pp-KTP source using an optical-fiber-based time-of-flight spectrometer that significantly outperforms previous techniques. We also present Hong-Ou-Mandel interference data showing that signal and idler photons are nearly indistinguishable. Finally, we describe second-order correlation measurements, which indicate that very little light is present in unwanted modes and that after becoming unentangled by interference on a beam splitter, the combinations of signal and idler behave as two independently squeezed vacua. We have observed up to 0.11 photons per pulse per mode for our highest pump pulse energy of 8.8 nJ, which corresponds to about 2.8 dB of squeezing in each mode. The measurements presented in this work are based on coincidence detection and are therefore insensitive to losses.

## 2. Background

During Type II down-conversion squeezed light is created in both vertical $V$ and horizontal $H$ polarizations, which we call the signal and idler beams. The signal and idler may be made of many entangled modes that overlap both in real space and frequency. Ideally the initial multi-mode state is pure, but photon loss will cause decoherence. By collecting the signal and idler with single mode optical fibers, we ensure that all light we detect exists in the same spatial mode. This spatial filtering may cause photon loss, but our measurements are insensitive to loss, so for simplicity we assume that the signal and idler are created in a single spatial mode. In the low squeezing limit, the initial, pure state of the signal and idler is then given by [16]:

$$|\xi\rangle_2 \simeq \sqrt{1-\xi}|0\rangle + \sqrt{\xi}\int d\omega_s \int d\omega_i \Psi(\omega_s,\omega_i) \hat{a}_V^\dagger(\omega_s) \hat{a}_H^\dagger(\omega_i)|0\rangle + \cdots, \quad (1)$$

where $\xi$ is the squeezing parameter, and the state is not normalized. $\Psi(\omega_s,\omega_i)$ is the signal-idler joint wavefunction, which describes the amplitude distribution for creating a photon pair with signal frequency $\omega_s$ and idler frequency $\omega_i$. This joint wavefunction is determined by both energy conservation and phase-matching conditions, and can be tailored by carefully designing the nonlinear crystal to account for the pump laser's spectral and spatial mode profiles. Note that higher order terms than those shown in the expansion above are relevant in our experiment, but they are not necessary for this definition of $\Psi(\omega_s,\omega_i)$.

We can describe a general squeezed state by expressing $\Psi(\omega_s,\omega_i)$ with a set of orthogonal modes using the Schmidt decomposition [8]:

$$\Psi(\omega_s,\omega_i) = \sum_n \sqrt{\lambda_n} \psi_n(\omega_s) \varphi_n(\omega_i), \quad (2)$$

where normalization requires $\sum_n \lambda_n = 1$. One measure of the quality of the squeezing is the effective mode number,

$$K = \left(\sum_n \lambda_n^2\right)^{-1}. \quad (3)$$

For all states $K \geq 1$. If only one nonzero coefficient is present (say $\lambda_1=1$), then the two-photon wavefunction is *factorizable*, $\Psi(\omega_s,\omega_i) = \psi_1(\omega_s)\phi_1(\omega_i)$, and $K = 1$. The effective mode number $K$ is a measure for the number of modes and is therefore not directly linked to the squeezing purity of any single mode (or pair of entangled modes) that can be obtained from homodyne measurements. Unfortunately, measurements of the joint spectral probability distribution, $|\Psi(\omega_s,\omega_i)|^2$, such as those described [11, 13-15] and in Section 3 below, do not give information about the phase of $\Psi(\omega_s,\omega_i)$. As a result, $K$ can only be computed from joint spectral probability measurements alone by making assumptions about this phase. Here, we perform the Schmidt decomposition of $|\Psi(\omega_s,\omega_i)|$ and report its effective mode number $K_{ABS}$. If we were to assume that the phase of $\Psi(\omega_s, \omega_i)$ were constant over all frequencies (or equivalently that $|\Psi(\omega_s,\omega_i)|$ were Fourier transform limited in frequency and time), then $K_{ABS}$ and $K$ would be identical. Without measuring the phase, we are unable to verify this assumption. While it is the case that all factorizable states must have $K = K_{ABS} = 1$, we note that measuring $K_{ABS} = 1$ does not necessarily guarantee that $K = 1$.

In addition to factorizability, we also hope to engineer the squeezing to produce identical frequency modes so that $\phi_1(\omega) = \psi_1(\omega)$, and hence $\Psi(\omega_s,\omega_i) = \psi_1(\omega_s)\psi_1(\omega_i)$. Meeting this condition will allow both signal and idler to be mode-matched to a single local oscillator field.

## 3. Measurement of the joint spectral probability distribution

### 3.1 Measurement techniques

Figure 1 shows a schematic of the experimental apparatus used to measure the individual signal and idler spectra, as well as the joint spectral probability distribution. Type II down-conversion in the pp-KTP crystal creates pairs consisting of one horizontally polarized signal photon (*H*) and one vertically polarized idler photon (*V*). To obtain an approximately circular joint spectral probability distribution with degenerate signal and idler photons, we calculated an optimum poling period of 46.55 μm and a crystal length of 2 mm. For a discussion of crystal engineering see [17]. To generate the local oscillator field for future homodyne measurements our experimental setup dictates a center pump wavelength of 785 nm, resulting in a down-conversion center wavelength of 1570 nm. This center wavelength is very close to the optimal down-conversion wavelength of 1582 nm, where the group index of the pump is equal to the mean of the signal and idler group indexes. A center wavelength of 1570 nm means the joint spectrum will be slightly non-circular, i.e. $\phi_1(\omega) \approx \psi_1(\omega)$. The crystal length was chosen to achieve transform limited signal and idler pairs based on the femtosecond pump's spectral bandwidth of 5.35 nm (FWHM), and to also match the spectral and temporal bandwidth of the local oscillator field that will be used in future studies.

The signal and idler photons are separated at a polarizing beam splitter and coupled into two individual telecom single-mode fibers. The *V* photons are delayed by approximately 180 ns before being recombined with the *H* photons at a 50/50 fiber beam splitter. This time-multiplexing allows the detection of both photons with only one single-photon detector, and the 180 ns delay circumvents the ~70 ns single-channel deadtime of the time-stamping electronics. Fewer than 5 % of pump pulses generate more than one signal-idler pair.

After recombination, both signal and idler photons are sent to the time-of-flight fiber spectrometer, which relies on the known dispersion of a long piece of optical fiber and an ungated single-photon detector with low timing jitter and low dark-count rate. Measuring the transit time of a photon through this 1.3 km-long fiber thus yields a measure of the photon wavelength. The detector is a NbTiN-based superconducting nanowire single-photon detector (SNSPD) packaged in a closed-cycle refrigerator [18]. A polarization scrambler before the detector eliminates artifacts caused by the polarization-dependent detection efficiency of the SNSPD [19] acting in combination with the wavelength-dependent polarization rotation in the optical fiber. The polarization scrambler reduces the overall system detection efficiency to about 4 % at 1550 nm, but assures a uniform detector response to all polarizations at each wavelength. The dark count rate of the SNSPD is about 1 kHz.

Each time the SNSPD detects a signal or idler photon, it delivers a voltage pulse to the time-tagging electronics. The arrival times are recorded with respect to a reference clock signal, which is provided by the fast photodiode triggered by the pump pulse shown in Fig. 1(a). The combined timing jitter of the detector and electronics is ~65 ps FWHM, and the jitter profile has a very clean, Gaussian shape, as evidenced by the red curve in Fig. 1(c) [20]. The jitter of the time tagging electronics alone is ~20 ps FWHM.

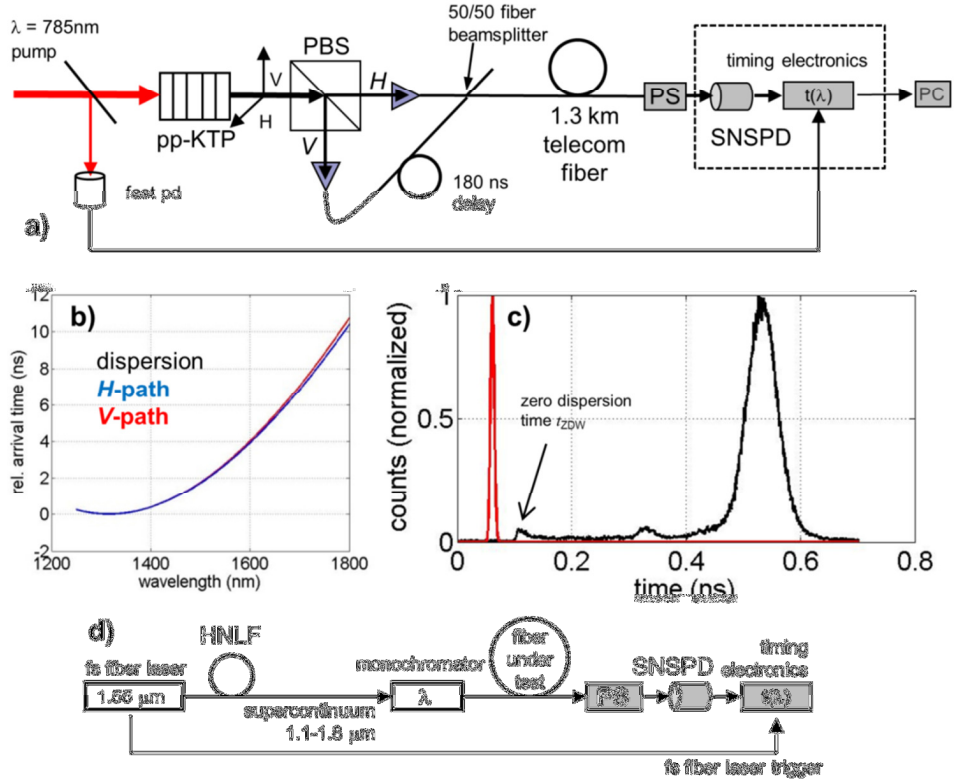

Fig. 1. (a) Experimental setup for measuring the joint spectral probability distribution output. pp-KTP = non-linear periodically poled KTP crystal; pd = photodiode; PBS = polarizing beam splitter; PS = polarization scrambler; SNSPD = superconducting nanowire single-photon detector; PC = computer for post-processing. (b) Measured dispersion curves of the fiber spectrometer *H* (blue line) and *V* (red line) paths. (c) Single-photon time-of-arrival histograms. The red line shows the instrument response curve (no dispersive fiber added). The black solid line shows single-photon (*H*) time-of-arrival after propagating through 1.3 km of single-mode telecom fiber. (d) Setup for dispersion calibration. HNLF = highly nonlinear fiber (supercontinuum source).

To determine the conversion factor between arrival time and wavelength, we calibrate the 1.3 km-long fiber by performing the group-delay measurement sketched in Fig. 1(d). A femtosecond fiber laser pumps a highly nonlinear fiber, creating a supercontinuum with a spectrum spanning from ~1.1 μm to ~1.8 μm. The supercontinuum is sent through an optical spectrum analyzer, which acts as a narrowband (~1 nm bandwidth) monochromator to select a particular test wavelength, which is in turn sent to the fiber spectrometer. The SNSPD and electronics record the time of arrival for photons at each wavelength. Arrival times are measured with and without the 1.3 km-long fiber in place; the difference of these two measurements yields the dispersion of the fiber. At 1570 nm, we measured a dispersion of 24.4 ± 0.2 ps·nm$^{-1}$ and 23.6 ± 0.2 ps·nm$^{-1}$ for the signal and idler path, respectively. Combined with our 1σ temporal resolution of ~28 ps (~65 ps FWHM), this results in 1σ wavelength measurement uncertainties of ~1.2 nm for both the signal and idler paths. In principle, increasing the length of the fiber should improve the spectral resolution; however, our ability to stabilize the fiber temperature, as well as the maximum dispersion set by the repetition rate of the laser will set a practical upper limit on fiber length. A third-order polynomial fitted to the data in Fig. 1(b) determines 1319 ± 0.3 nm as the zero-dispersion-wavelength of this fiber. Determining the spectrum of the signal or idler alone involves acquiring a time-domain histogram proportional to the distribution of arrival times of signal or idler photons with respect to the reference clock signal. An example idler timing-spectrum is shown in Fig. 1(c): in addition to the main parametric down-conversion peak centered close to 1570 nm, there is a weak, broad spectral output, with some detectable light at wavelengths shorter than the zero-dispersion wavelength of 1319 nm. Photons at 1319 nm are the first to arrive at the detector; any photon with a shorter or longer wavelength arrives later, and the spectrum is folded over about this point. The resulting sharp edge in the time-resolved data, identified by the arrow in Fig. 1(c), gives a reference for the absolute wavelength calibration, because it pinpoints the time at which 1319 nm photons arrive at the detector. To measure the joint spectral probability distribution, time-tagged data are post-processed to generate two-dimensional coincidence histograms as a function of both signal and idler wavelengths. Our technique is similar to an earlier approach [14] that used InGaAs avalanche photodiodes that required temporal gating due to the high probability of dark counts and afterpulsing. The time gates were adjusted sequentially to acquire one specific signal-idler wavelength pair at a time [14]; as a result, this prior work did not realize the full power of this technique. Here, by contrast, our SNSPDs can operate in free-running mode, allowing us to acquire coincidences for all signal-idler wavelength pairs in parallel, enabling an $N^2$ reduction in measurement time for an $N \times N$ array of signal and idler wavelengths.

### 3.2 Joint spectral probability dependence on pump geometry

Figures 2(a)-(c) show the measured joint spectral probability distributions for three different pump waists inside the crystal. The confocal parameters, $b_c$, (in air) of the focused beams are 0.25 mm, 5.0 mm and 64 mm. These correspond to pump waists $w_0$ of 8 μm, 35 μm and 126 μm, respectively. The insets in Fig. 2(a)-(c) illustrate how the pump beam size changes as it propagates through the crystal for each waist size. All joint spectra were taken with approximately equal measurement times (~4 hours). For the smallest pump waist (Fig. 2(a)), the joint spectrum is elliptical and tilted and therefore is clearly not factorizable. Figures 2(b) and 2(c) show approximately factorizable spectral distributions. However, at the largest pump waist, the photon collection efficiency into the single-mode fiber drops significantly and the data clearly become noisier. This is in qualitative agreement with theoretical results reported by R. Bennink et al. [21]. The sinc-function side lobes originating from the phase-matching condition are visible in the joint spectral output. These side lobes lead to a lower factorizability of the produced state and will need to be addressed by spectral filtering or apodizing the poling period for future use in generating high-fidelity optical cat states [22]. The plots in the lower half of Fig. 2 show the individual spectra of H-polarized signal (blue) and V-polarized idler (red) photons. All single signal and idler spectra were acquired by measuring the arrival times of the signal and idler photons alone, separate from the joint spectral distribution measurement. Because this measurement does not require coincidence events between signal and idler photons, these data have much smaller uncertainties. The signal and idler spectra are approximately mirror images of one another, reflected about a line of symmetry defined by the pump degeneracy point, which is 1570 nm.

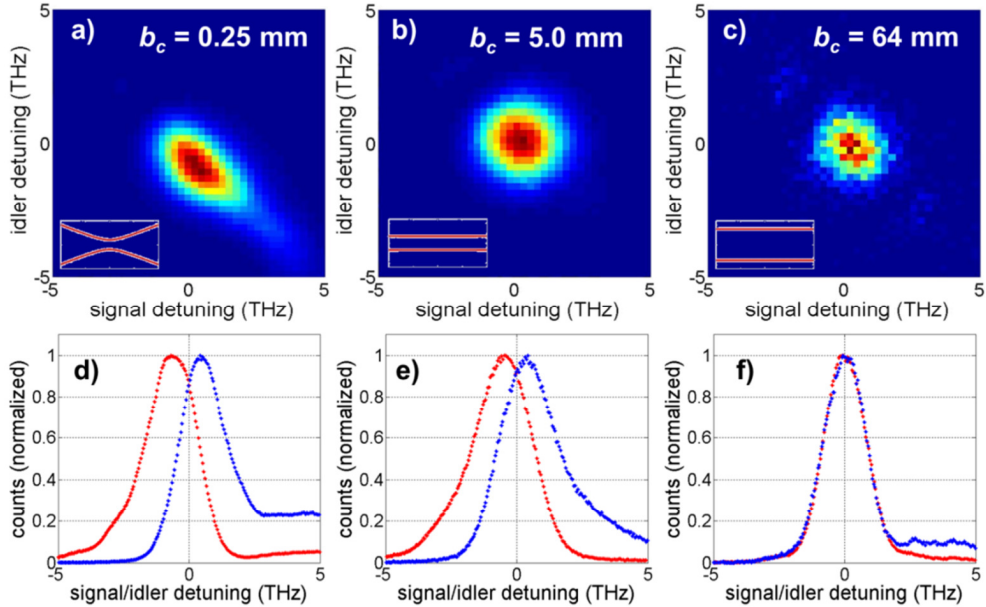

Fig. 2. (a)-(c) Joint spectral probability distributions for three different pump beam waists. The confocal parameter ($b_c$) of each pump beam is listed in the upper figure, and the relative size and divergence of each pump beam is drawn in the lower left of each joint spectrum. The edges of the insets mark the region around the beam waist inside the crystal. The single-photon signal (blue) and idler (red) spectra are shown in (d), (e) and (f). These are measured under the same conditions as the joint spectra in (a), (b) and (c), respectively. The 1σ uncertainty in frequency is ~0.14 THz. This is derived from the photon arrival time uncertainty and corresponds to a 1σ wavelength uncertainty of ~1.2 nm. The vertical uncertainty can be identified by the scatter in the data and is dominated by the uncertainty of the photon count statistics. Note that (a), (b), and (c) have different color scales, and (d), (e), and (f) use different normalizations. The 10 THz range of the x-axis corresponds to a wavelength range of ~82 nm.

This mirroring, which is somewhat diminished due to the reduced SNSPD detection efficiency at longer wavelengths (lower frequencies) [18], is due to energy conservation in the down-conversion process. A slight non-degeneracy of H and V photons can be seen from the separation of the individual peaks. For the smallest pump waist we observe a separation of ~6 nm. This decreases to ~4 nm for the intermediate beam waist, and approximately degenerate spectra are achieved for the largest pump waist, demonstrating $|\phi_1(\omega)|^2 \approx |\psi_1(\omega)|^2$. A possible reason for the shift from the degeneracy point to non-degenerate pairs is the contribution of higher-order spatial modes that overlap with the spatial mode of our single-mode fiber. A more uniform pump waist inside the crystal thus yields a higher degree of degeneracy, but at the expense of reduced pump fluence and reduced fiber coupling efficiency of the photon pairs.

*3.3 The effect of spectral filtering on the joint spectral probability density and Schmidt decomposition*

Next, to be able to also use the high quantum efficiency transition edge sensor, we switch from directly using the 76 MHz output of the Ti:Sapphire oscillator to using 456 kHz, cavity-dumped pulses. The top two panes in Fig. 3 show (a) the absolute value of the joint spectral amplitude distribution $|\Psi(\omega_s,\omega_i)|$, calculated from the measured joint-probability distribution, and (b) the individual signal and idler spectra (measured by determining the single photons' arrival times), all for a pump waist of ~50 μm ($b_c$ = 10 mm) and without spectral filtering. The maxima of the signal and idler spectra are separated by approximately 2 nm (determined from the peak separation of the two Gaussian fits): the spectral bandwidth and center wavelengths are 20.1 nm (20.5 nm) and 1569 nm (1571 nm) for the H (V) photons. The spectral decomposition of the absolute value of the joint spectral amplitude distribution yields $K_{ABS}$ = 1.06 ± 0.02. The uncertainties for all the reported effective mode numbers were found by parametric bootstrap resampling [23]. The data presented in Fig. 3(c) & (d) were recorded while the output of the source was filtered with a top-hat spectral filter with a center wavelength of 1570 nm and a bandwidth of 8.6 nm (1.05 THz).

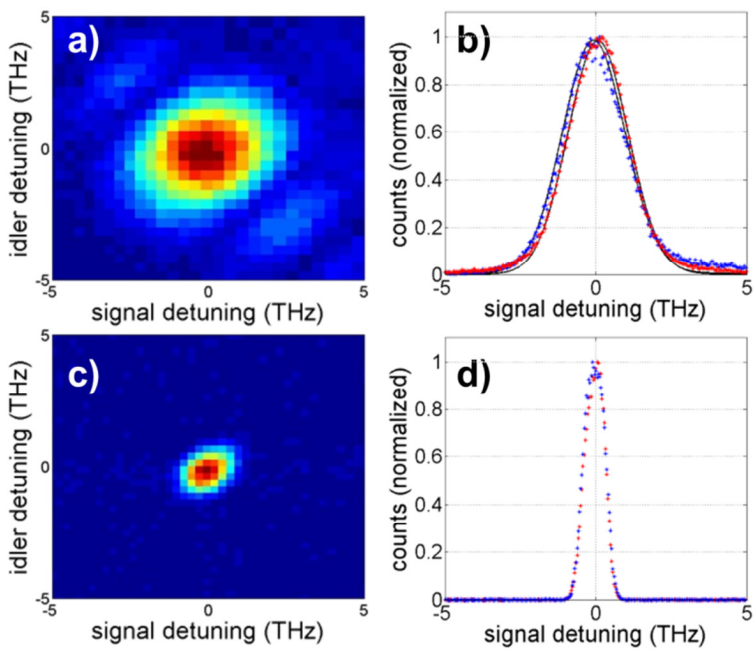

Fig. 3. (a) Absolute value of the joint spectral amplitude distribution for squeezing generated using cavity dumped pulses. This was calculated from the measured joint spectral probability distribution. (b) Corresponding individual signal (blue) and idler (red) spectra. The dashed lines are Gaussian fits to the data. (c) Absolute value of the joint spectral amplitude distribution calculated from the measured joint spectral probability distribution for the spectrally filtered source. (d) Individual signal (blue) and idler (red) spectra for the spectrally filtered source. Note that (a) and (c) have different color scales. The vertical uncertainties in (b) and (d) can be identified by the scatter in the data and is dominated by the uncertainty of the photon count statistics. The 1σ frequency uncertainty is ~0.14 THz derived from the photon arrival time uncertainty.

The decomposition of $|\Psi(\omega_s,\omega_i)|$ yields a slightly larger $K_{ABS}=1.08 \pm 0.02$ than in the unfiltered case, but the signal and idler are closer to being fully degenerate. One can observe a slightly elliptical output joint spectrum in both cases, which was caused by a change in the bandwidth of the pump laser in the cavity-dumped configuration [14]. Applying the measured filter response to the main modes of the spectral decomposition from the unfiltered dataset shown in Fig. 3(a), we calculate a signal and idler mode transmission of ~30 % and ~32 %, respectively. All other higher order modes have a total transmission of less than 5 % and 4 % for the vertically and horizontally polarized modes, respectively. This filter is not optimally designed to pass the main modes only. To compare the overall transmission of the higher order modes through the filter used in the experiment, we calculated the bandwidth of a perfect top-hat filter which will pass less than 5 % of the higher order modes. We found that this filter has a bandwidth of 0.7 THz (5.8 nm) and will allow 50 % and 54 % of the signal and idler main mode to pass. We also calculated the bandwidth of a top-hat filter that passes more than 99 % of the signal and idler main modes; this bandwidth is 3.4 THz (28 nm) and this filter would transmit less than 65 % of the higher order modes. If the latter filter was used, less than 0.1 % of all transmitted photons would be in unwanted modes. We will limit the remainder of this paper to measurements taken with the 8.6 nm bandwidth spectral filter placed after the down-conversion source.

## 4. Hong-Ou-Mandel interference

To quantify the indistinguishability of signal and idler photons, we measured the Hong-Ou-Mandel (HOM) interference between them. Figure 4(a) shows the experimental scheme. After signal and idler are generated and spectrally filtered, they are split at a polarizing beam splitter. The polarization of the $V$ mode is rotated by 90° with a half-waveplate ($\lambda/2$), and signal and idler are then recombined on a 50/50 beam splitter. A computer-controlled translation stage varies the arrival time ($\Delta t$) at the beam splitter of photons in mode $H$. Each output mode ($c$ and $d$) of the 50/50 beam splitter is coupled into a single-mode fiber and the collected photons are delivered to a photon-number-resolving, superconducting transition-edge sensor (TES) [24] for each output mode. The high efficiency of the TES detectors ensures detection of 95 % and 73 % of the photons that are coupled into the fibers in modes $c$ and $d$, respectively. (The reason for the lower detection efficiency of one detector is that this detector was optimized for a wavelength band around 800 nm, yielding a lower efficiency at 1550 nm).

For each pump pulse, each TES records the number of photons ($n$ = 0, 1, 2,...) in each mode. A coincidence is an event in which both detectors record $n \geq 1$ photons for the same pump pulse. These coincidences are recorded as a function of $\Delta t$. The increased timing jitter (~100 ns) of the TES dictates that we perform this experiment with the cavity-dumped pulse train, at a repetition rate of 456 kHz.

If the squeezing source were weakly pumped and only produced signal-idler photon pairs, and if the spatio-temporal modes of the signal and idler photons were identical, then coincident clicks of TES1 and TES2 would be impossible at zero delay. However, some zero delay coincidences will be produced by background photons and (because we use strong pumping) by events in which multiple photon pairs are produced. The HOM interference data are presented in Fig. 4(b) for pump pulse energy of 1.4 nJ just before the crystal. The red datapoints shows the raw data, which has an HOM interference visibility of 86 ± 3 %.

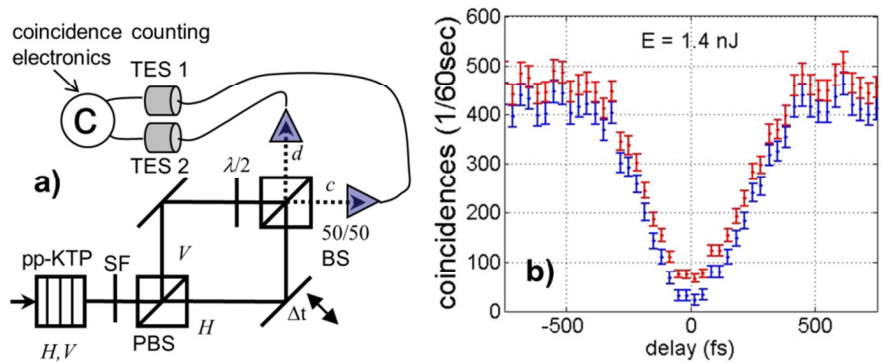

Fig. 4. (a) Experimental geometry for measuring Hong-Ou-Mandel interference between signal and idler photon. SF = spectral filter with $\lambda$ = 1570 and $\Delta\lambda_{FWHM}$ = 8.6 nm. (b) HOM results with (blue) and without (red) correcting for background photon-coincidence events.

We calculate the HOM interference visibility by calculating the ratio of the minimum intensity at the dip and the average maximum intensity. The uncertainty is determined via propagation of the counting statistics uncertainties. When we block the pump laser, the background coincidence rate can be determined. These background coincidences arise when both TES detectors simultaneously register counts due to stray light or blackbody photons. After subtracting these accidental background counts the HOM interference visibility increases to 90 ± 3 %. We can also measure the single signal/idler photon and single background count rate on each detector and calculate the accidental coincidence rate between one down-converted photon at one detector and one background photon at the other detector. When also correcting for these accidental coincidences, the HOM interference visibility is further increased to 95 ± 4 % (blue datapoints). High visibility HOM interference does not necessarily imply high purity squeezing. An increase of the effective mode number $K$ would likely cause a decrease of the HOM interference visibility, because not all signal and idler mode pairs would be fully degenerate. We compare the visibility of our HOM interference with the overlap integral of the measured signal and idler photons' spectra shown in Fig. 3(d) [13]: $\left|\int |\Psi(\omega)| |\varphi(\omega)| d\omega\right|^2 \approx 98$ %. (Note that this is only an estimate. A full calculation of the visibility would require knowledge of the joint spectral amplitude, but we only know its absolute value. Also, the full calculation should consider entanglement between the signal and idler photons' wave functions, which we have neglected here.) This value roughly agrees with the corrected HOM visibility. Multi-pair generation in the pp-KTP crystal is also expected to further reduce the HOM visibility, since the HOM effect predicts full coincidence suppression only when exactly one photon is present at each input of the 50/50 beam splitter [25, 26]. However, for the data in Fig. 5(b), multi-pair generation does not contribute a significant number of extra coincidences (~1 % of the single-pair event rate), and we therefore did not correct for this additional contribution. Nevertheless, multi-pair events are not undesirable for pure squeezed light generation. In fact, we strive to create a source with high squeezing, which would entail a high probability of multi-pair generation.

## 5. Second-order correlation measurements

In the final phase of this experiment, we use second-order correlation measurements to verify (1) that the signal and idler modes contain entangled squeezed modes; (2) that the signal and idler modes can be dis-entangled by interference on a beam splitter, and (3) that very little light is present in unwanted modes. The second-order correlation at zero delay between modes $j$ and $k$ at zero time delay is defined to be

$$g^{(2)}_{jk} = \frac{\langle \hat{a}^\dagger_j \hat{a}^\dagger_k \hat{a}_j \hat{a}_k \rangle}{\langle \hat{a}^\dagger_j \hat{a}_j \rangle \langle \hat{a}^\dagger_k \hat{a}_k \rangle}. \quad (4)$$

When $j=k$, we refer to the $g^{(2)}_{ij}$ as the autocorrelation, and for $j \neq k$ it is the cross-correlation. $g^{(2)}$ is insensitive to linear photon loss (provided that all modes are subjected to the same loss), so it can be a useful diagnostic tool regardless of the efficiency of the photon detection systems, but it is unable to distinguish between a pure squeezed state and the same squeezed state subject to loss (or a similar effect).

The first method for measuring $g^{(2)}$ uses a standard photon-correlation scheme with two SNSPDs and time-domain histogramming electronics. A photon detected by one SNSPD starts a timer, which is then stopped after a time delay $\tau$, once the other SNSPD detects a photon. The resulting histogram of start-stop pairs is proportional to $g^{(2)}(\tau)$. Figure 5(a) shows an example time-domain histogram. To determine the zero delay value, $g^{(2)}$, we take the area of the peak at zero delay (identified by the large peak in figure 5(a)) and divide by the average area of the surrounding peaks. To find the autocorrelation of a single mode, $g^{(2)}_{jj}$, we place a 50/50 beam splitter in mode $j$, and place one SNSPD at each output port of the beam splitter. The configuration for measuring $g^{(2)}_{HH}$ in this way is shown in Fig. 5(e). To find the cross-correlation between two different modes, $g^{(2)}_{jk}$ for $j \neq k$, we place one SNSPD in mode $j$ and the other in mode $k$, as shown in Fig. 5(f) for $g^{(2)}_{HV}$. When measuring $g^{(2)}_{jk}$ by time-domain histogram we use the direct 76 MHz output of the Ti:Sapphire oscillator.

The second method for determining the second-order auto-correlation uses photon-number probabilities measured with a TES detector and the expression [27]:

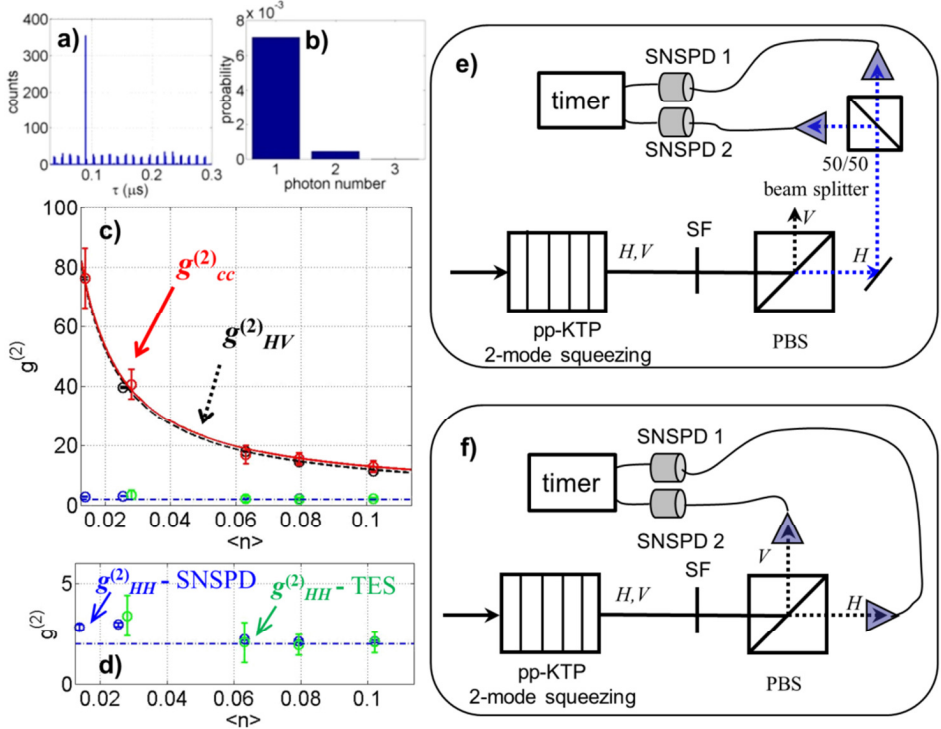

Fig. 5. Second-order correlation experiment. (a) Example of time-domain histogram measured by SNSPDs. (b) Example photon-number probabilities measured by TES for a squeezed vacuum input state. (c) $g^{(2)}$ results. The red circles show single-mode squeezed vacuum $g_{cc}^{(2)}$ data, and the red solid line is a plot of Eq. (8). The black circles show cross-correlation data $g_{HV}^{(2)}$, and the black dashed line is the plot of Eq. (6). The blue circles show thermal-state data $g_{HH}^{(2)}$ measured with time-domain histogramming by SNSPDs, the green open circles show $g_{HH}^{(2)}$ measured with photon-number probabilities from a TES, and the blue dashed line is the plot of Eq. (7). All error bars were calculated by linear propagation of the photon count uncertainties assuming poissonian distribution of the photon count statistics. Due to much longer measurement times, i.e., many more photon counts, when we used the time histogramming method, the error bars for $g_{HV}^{(2)}$ (black circles) and $g_{HH}^{(2)}$ (blue circles) are within the size of the markers. To estimate $\langle n \rangle$ and plot Eq. (6) and (8) we estimated the proportionality constant between pulse energy and generated mean photon number. The pump pulse energy for all data ranged from 1.2 nJ to 8.8 nJ (d) Magnified view of $g_{HH}^{(2)}$. (e) Experimental scheme for measuring $g_{HH}^{(2)}$. (f) Experimental scheme for measuring $g_{HV}^{(2)}$.

$$g_{jj}^{(2)} = \frac{\sum_{n=0}^{\infty} n(n-1) p_n}{\left(\sum_{n=0}^{\infty} n p_n\right)^2} = \frac{2 p_2 + 6 p_3 + 12 p_4 + ...}{(p_1 + 2 p_2 + 3 p_3 + 4 p_4 + ...)^2}, \quad (5)$$

where $p_n$ represents the probability of detecting $n$ photons from any given pump pulse. (We do not use this method for cross-correlation.) Figure 5(b) shows an example photon-number probability distribution. For clarity, we have not plotted the vacuum contribution to this distribution, but it can be easily computed as $p_0 = 1 - \sum_{n=1}^{\infty} p_n$. Due to the slow response of the TES, when we measure $g_{jj}^{(2)}$ via photon-number probabilities, we operate with the cavity dumped output at 360 kHz.

For the cross-correlation between the H and V modes of an ideal two-mode squeezer, we expect [28]:

$$g_{HV}^{(2)} = 2 + \frac{1}{\langle n \rangle}, \quad (6)$$

where $\langle n \rangle$ is the mean number of photons generated per pump pulse. If multiple modes are present, $g_{HH}^{(2)}$ is expected to be smaller, approaching 1 as the number of modes approaches infinity.

A diagram of this measurement is in Fig. 5(f). The result of this measurement is shown for several ⟨n⟩ values in Fig. 5(c) with black circles. The black curve is a plot of Eq. (6) to this data. Note that we estimated the proportionality constant between pulse energy and generated mean photon number. We used this proportionality constant to plot Eq. (6) and (8) (red and black lines in Fig. 5)

The autocorrelation of each individual mode should behave as a thermal source [1],

$$g_{HH}^{(2)} = g_{VV}^{(2)} = 2, \qquad (7)$$

independent of ⟨n⟩. Values for $g_{HH}^{(2)}$ obtained using both time-domain histograms measured by SNSPD (as shown in Fig. 5(e)) and photon-number probabilities measured by TES appear in Fig. 5(c) and (d). The results from both methods are in close agreement. We would expect the presence of multiple squeezed modes entering the detectors to decrease the value of $g_{HH}^{(2)}$ from 2 to 1+1/K, and this property was used to estimate K in [11]. However, we find $g_{HH}^{(2)}$ to be somewhat higher than 2 at the lowest pump powers; this is likely due to experimental imperfections that cause some V photons to scatter into the H output port of the polarizing beam splitter. This problem is worse at small ⟨n⟩ simply because the bunching between H and V is much more pronounced as ⟨n⟩ decreases, as is evident in the $g_{HV}^{(2)}$ data.

After H and V experience HOM interference as in Fig. 4(a), we expect a single-mode squeezed vacuum in each output port of the 50/50 beam splitter [29], each of which should have auto-correlation given by

$$g_{cc}^{(2)} = g_{dd}^{(2)} = 3 + \frac{1}{\langle n \rangle}. \qquad (8)$$

Results for $g_{cc}^{(2)}$ obtained using the TES measurement of photon probabilities are shown in Fig. 5(c) with red circles, and the red line is the plot of Eq. (8). Again we see good agreement with expectations.

## 6. Summary

We have created a periodically poled KTP squeezing source that delivers circular joint spectral probability distributions. When applying spectral filtering with a bandwidth of 8.6 nm, we observe a Hong-Ou-Mandel interference visibility of 95 % when correcting for accidental counts. The spectral filter helps in matching the signal and idler modes and in suppressing unwanted modes, but it does cause loss to the dominant squeezed mode. Careful engineering of the filter might allow the dominant mode to pass with high transmissivity while also achieving our mode matching goals. The measured $g^{(2)}$ values fit the theoretical predictions of single-mode outputs (thermal and squeezed vacuum) very well. All of these results give evidence that we have successfully created entangled squeezed signal and idler beams with nearly identical, factorizable, spatio-temporal modes. They also show that very little squeezed light is created in unwanted modes that are also collected by our photon detectors. We have employed advanced photon-counting techniques based on superconducting detectors (SNSPDs and TESs) to characterize the single-mode character of two-mode and single-mode squeezed states. These techniques are useful tools for investigating the spatial-mode properties of the squeezing. In the near future, we plan to map the local oscillator mode-matching characteristics and the squeezing purity of this source using homodyne detection.

## Acknowledgements

This work was supported by the NIST 'Innovations in Measurement Science' Program and the Quantum Information Science Initiative (QISI). RHH gratefully acknowledges support from the UK Engineering and Physical Sciences Research Council and a Royal Society University Research Fellowship. VZ and SND acknowledge NWO (VIDI grant).